# Highly Luminescent Bulk Quantum Materials Based on Zero-Dimensional Organic Tin Halide Perovskites

Chenkun Zhou[1], Zhao Yuan[1], Yu Tian[2], Haoran Lin[1], Ronald Clark[3], Banghao Chen[3], Lambertus J. van de Burgt[3], Jamie C. Wang[3], Kenneth Hanson[2,3], Quinton J. Meisner[3], Jennifer Neu[4], Tiglet Besara[4], Theo Siegrist[1,2,4], Eric Lambers[5], Peter Djurovich[6], Biwu Ma[1,2,3*]

A crystalline solid is a material whose constituents, such as atoms, molecules or ions, are arranged in an ordered microscopic structure, forming a periodic lattice that extends in all directions. The interactions between the lattice points could lead to the formation of band structure.[1] As a result, the properties of typical inorganic crystals show strong dependence on their size, especially in nanoscale, or so called quantum size effect.[2] The molecular interactions in organic crystals cause their properties distinct from those of individual molecules.[3] Single crystalline materials that exhibit bulk properties consistent with their individual building blocks, or bulk assemblies of quantum confined materials without band formation or quantum size effect (bulk quantum materials), are rare to our best knowledge. Here we report a class of bulk quantum materials based on zero-dimensional (0D) tin halide perovskites $((C_4N_2H_{14}X)_4SnX_6, X = Br, I)$, in which the individual tin halide octahedrons $(SnX_6^{4-})$ are completely isolated from each other and surrounded by the organic ligands

[1]Chemical and Biomedical Engineering, FAMU-FSU College of Engineering, Florida State University, Tallahassee, FL 32310, USA. [2]Materials Science and Engineering, Florida State University, Tallahassee, FL 32306, USA. [3]Department of Chemistry and Biochemistry, Florida State University, Tallahassee, FL 32306, USA. [4]National High Magnetic Field Laboratory, Florida State University, Tallahassee, FL 32310, USA. [5]Research Service Centers, University of Florida, Gainesville, Florida 32661, USA. [6]Department of Chemistry, University of Southern California, Los Angeles, California 90089, USA

**($C_4N_2H_{14}X^+$). The complete site isolation of the photoactive $SnX_6^{4-}$ by wide band gap $C_4N_2H_{14}X^+$ leads to strong quantum confinement without band formation between $SnX_6^{4-}$, allowing the bulk crystals to exhibit the intrinsic properties of the individual metal halide octahedrons. These 0D perovskites can also be considered as perfect host-dopant systems[4], with photoactive $SnX_6^{4-}$ periodically doped in the wide band gap matrix composed of $C_4N_2H_{14}X^+$. Highly luminescent strongly Stokes shifted broadband emissions with photoluminescence quantum efficiencies (PLQEs) of up to near-unity were realized, as a result of excited state structural reorganization of individual $SnX_6^{4-}$ in the quantum confined 0D structure. Our discovery of bulk quantum materials that exhibit bulk properties consistent with their individual building blocks opens up a new paradigm in functional materials design.**

Hybrid organic-inorganic metal halide perovskites, consisting of a wide range of organic cations and inorganic anions, are an important class of crystalline materials with exceptional structural tunability. By choosing appropriate organic and inorganic components, the crystallographic structures can be finely controlled with the inorganic metal halide octahedrons forming zero- (0D), one- (1D), two- (2D), and three-dimensional (3D) structures surrounded by the organic moieties.[5,6] The rich chemistry of metal halide perovskites enables numerous ways of band gap control and color tuning. Highly luminescent 2D, quasi-2D, and 3D perovskites have been obtained with tunable narrow emissions, by controlling the chemical composition and quantum confinement.[7-10] Broadband emissions across the entire visible spectrum have also been realized in corrugated-2D and 1D perovskites, as a result of efficient exciton self-trapping in quantum well and quantum wire structures, respectively.[11-14] The excellent color tunability and high PLQE make metal halide perovskites highly promising light-emitting materials. However, the use of toxic heavy metal like lead represents a critical challenge for the potential wide adoption of these materials. All lead-free metal halide perovskites discovered to date have shown extremely low PLQEs.[15,16] In the present work, we have synthesized and characterized 0D lead-free tin halide perovkites (($C_4N_2H_{14}X)_4SnX_6$, X = Br, I) that exhibit Gaussian-shaped and strongly Stokes shifted broadband emissions with PLQEs of up to near-unity. Unlike previously reported 0D perovskites, such as $(CH_3NH_3)_4PbI_6 \cdot 2H_2O$ and $Cs_4PbBr_6$, showing extremely low luminescence and stability under ambient conditions[17], these highly luminescent Sn based 0D perovskites have great stability in air.

0D Sn halide perovskite crystals in millimeter size were prepared in high yield (> 70 %), by slowly diffusing dichloromethane into a precursor solution of Sn halide ($SnX_2$, X = Br, I) and N, N'-dimethylethylene-1,2- diammonium halide salt ($CH_3NH_2^+CH_2CH_2NH_2^+CH_3 \cdot 2X^-$) at room temperature in a $N_2$ filled glove box. The crystal structures of the Sn halide perovskites were determined using single crystal X-Ray Diffraction (SCXRD) (Extended Data Table 1), which show a 0D structure with individual Sn halide octahedral $SnX_6^{4-}$ ions completely isolated from each other and surrounded by $C_4N_2H_{14}X^+$ ions (Fig. 1A). A clear full overage of organic ligands on an individual Sn halide octahedron can be seen in a space-filling model in Fig. 1B. The complete site isolation between the photoactive metal halide octahedrons ($SnX_6^{4-}$) by the wide band gap organic species ($C_4N_2H_{14}X^+$), with the distance between two metal centers of > 1 nm as shown in Fig. 1C, leads to no interactions between $SnX_6^{4-}$ or band formation. Therefore, the potential energy diagram for this perfect host-dopant structure, or bulk assembly of core-shell quantum confined material, can be described as in Fig. 1D, which would enable the bulk crystals to exhibit the intrinsic photophyiscal properties of the individual tin halide octahedrons. The powder X-ray diffraction (PXRD) patterns of ball-milled crystal powders display exactly the same features as simulated patterns from SCXRD, suggesting a uniform crystal structure of as-synthesized bulk perovskite crystals (Extended Data Fig. 1). Solid state $^{119}$Sn nuclear magnetic resonance spectroscopy (NMR) was also used to characterize the structure of these 0D Sn halide perovskites. Only one Sn site at -546.8 and -382.2 ppm was observed for 0D $(C_4N_2H_{14}Br)_4SnBr_6$ and $(C_4N_2H_{14}I)_4SnI_6$, respectively (Extended Data Fig. 2), which is in agreement with the crystal structure and shielding features. Elemental analysis also confirmed the purity and uniformity of these 0D Sn halide perovskite crystals. To further verify the structure, composition, and only Sn (II) present in these 0D perovskites, we have performed X-ray photoelectron spectroscopy (XPS) measurements (Extended Data Fig. 3).

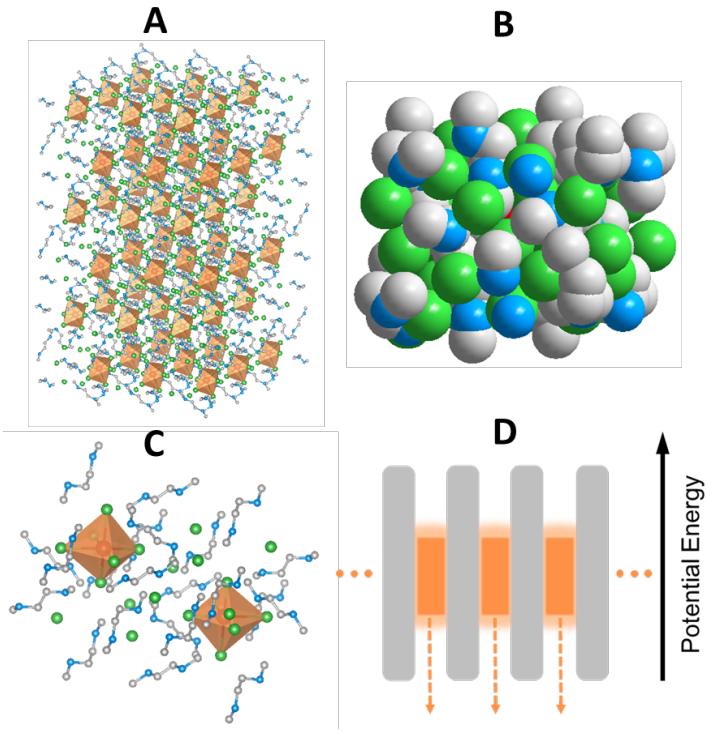

**Fig. 1. Single crystal structure and energy diagram of 0D Sn halide perovskites.** (**A**) View of the structure of 0D Sn bromide perovskite $(C_4N_2H_{14}Br)_4SnBr_6$ (red spheres: tin atoms; green spheres: bromine atoms; blue spheres: nitrogen atoms; gray spheres: carbon atoms; yellow polyhedrons: $SnBr_6^{4-}$ octahedrons; hydrogen atoms were hidden for clarity). (**B**) Space filling model with an individual $SnBr_6^{4-}$ completely covered by $C_4N_2H_{14}Br^+$ ions. (**C**). View of two unit cells with individual Sn bromide octahedrons $(SnBr_6^{4-})$ completely isolated from each other and surrounded by $C_4N_2H_{14}Br^+$ ions. (**D**) Schematic potential energy diagram of a host-dopant system $(C_4H_{14}N_2X^+\text{-}SnX_6^{4-})$.

The photophysical properties of 0D Sn halide perovskite crystals were characterized using UV-Vis absorption spectroscopy, as well as steady state and time-resolved photoluminescence spectroscopies. Major photophysical properties are summarized in Table 1. Fig. 2A shows the images of the bulk crystals under ambient light and a UV lamp irradiation (365 nm). Strong yellow and orange emissions under UV irradiation were observed for 0D $(C_4N_2H_{14}Br)_4SnBr_6$ and $(C_4N_2H_{14}I)_4SnI_6$, respectively. Fig 2B shows the excitation and emission spectra of the Sn halide perovskite crystals. The excitation maxima shift from 355 nm to 410 nm upon substitution of Br with I in the $SnX_6^{4-}$ octahedron, consistent with the weaker ligand field effect of I versus Br. The absorption spectra of these 0D perovskite crystals match well with their excitation spectra, except the scattering in the low energy regions (Extended Data Fig. 4). Color tuning by halide substitution has been extensively studied for 3D and 2D metal halide perovskites, which displays the same trend.[7,8] Emissions from 3D and 2D metal halide perovskites with band formation typically have small Stokes shifts and narrow full width at half maximum (FWHM) with nanosecond exciton lifetimes, due to the delocalized excitonic character of the excited states. In contrast, extremely large Stokes shifts (~ 200 nm) and FWHM (> 100 nm) are observed for these 0D Sn halide perovskites, which are similar to what observed in rare-earth doped phosphors with localized excited states.[18] To verify these emissions presenting the intrinsic properties of bulk crystals, we have measured the dependence of emission intensity on excitation power at room temperature. As shown in Fig. 2C, the intensity of the broadband yellow emission from the 0D $(C_4N_2H_{14}Br)_4SnBr_6$ exhibits a linear dependence on the excitation power up to 500 W/cm$^2$, suggesting that the emission does not arise from permanent defects.[11] The emissions of these crystals become narrower at 77K with FWHMs of ~ 63 nm (Fig. 2D), which is likely attributed to reduced thermally populated vibrational states at low temperature. The decay curves of broadband emissions from the bulk perovskite crystals at room temperature and 77K are shown in Fig. 2E, giving long lifetimes of ~ 2.2 μs for 0D $(C_4N_2H_{14}Br)_4SnBr_6$ and ~1.1 μs for 0D $(C_4N_2H_{14}I)_4SnI_{6,}$. The similar decay behaviors at room temperature and 77 K suggest little-to-no change of the characteristics of radiative and non-radiative processes. Unlike most bulk metal halide perovskites suffering from low PLQEs as compared to their nanocrystalline counterparts,[19] these bulk 0D perovskite crystals possess extremely high PLQEs at room temperature: 95 ± 5 % for $(C_4N_2H_{14}Br)_4SnBr_6$ and 75 ± 4 % for $(C_4N_2H_{14}I)_4SnI_6$ (Extended Data Fig. 5). It is noteworthy that these are highest PLQEs ever achieved for any lead-free metal halide perovskites. These Sn based materials also showed great stability under continuous high power mercury lamp irradiation (150 mW/cm$^2$) (Extended Data Fig. 6), as well as high thermal stability (Extended Data Fig. 7). Thermogravimetric analysis (TGA) shows that these materials do not decompose until 200 °C (Extended Data Fig. 8). This high stability in air is not surprising if we consider the unique core-shell structure having the photoactive Sn halide octahedrons well protected by the organic shells.

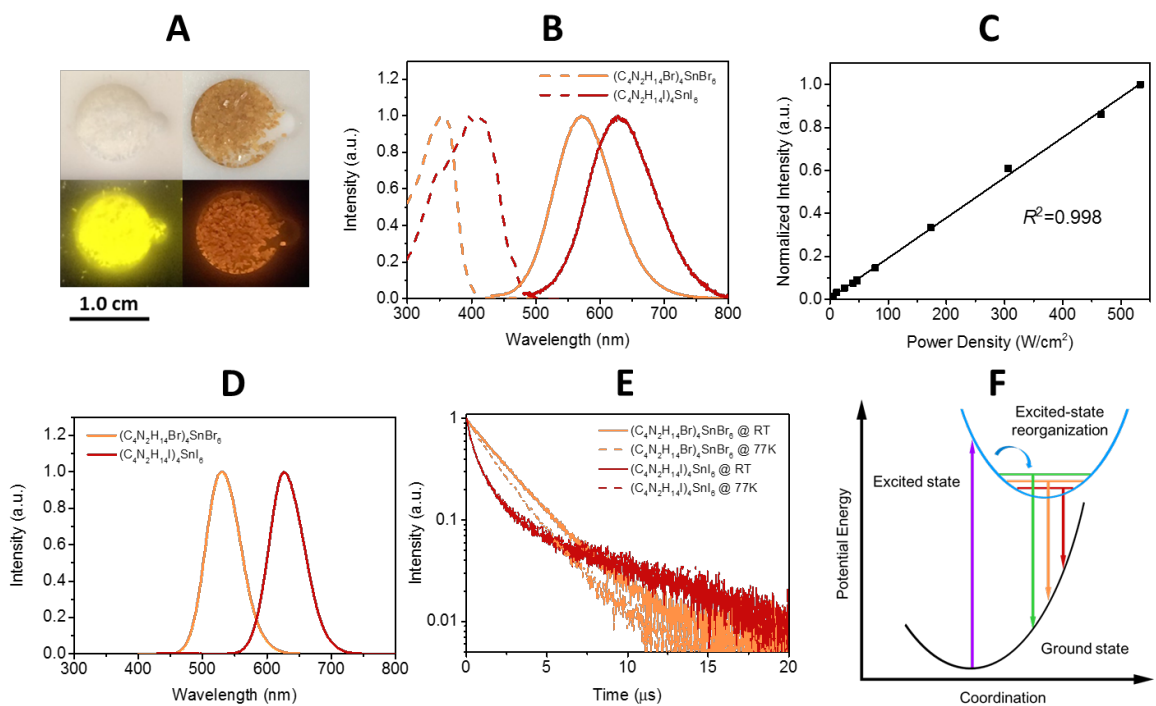

**Fig. 2. Photophysical properties of 0D Sn halide perovskites at room temperature and 77 K.** (**A**) Images of bulk 0D Sn halide perovskite crystals under ambient light and a UV lamp irradiation (365 nm). (**B**) Excitation (dash lines) and emission (solid lines) (excited at 360 nm) spectra of bulk 0D Sn halide perovskite crystals at room temperature. (**C**) Emission intensity versus excitation power for 0D $(C_4N_2H_{14}Br)_4SnBr_6$ at room temperature. (**D**) Emission spectra (excited at 360 nm) of 0D Sn halide perovskite crystals at 77 K. (**E**) The emission decays of 0D Sn halide perovskite crystals at room temperature and 77 K. (**F**) The mechanism of excited state structural reorganization: the straight and curved arrows represent optical and relaxation transitions, respectively.

**Table 1.** Photophysical properties of 0D Sn halide perovskites at room temperature and 77 K.

| 0D Perovskites | $\lambda_{exc}$, nm | $\lambda_{em}$, nm | FWHM, nm | Stokes shift, nm | $\Phi$, % | $\tau_{av}$, µs |
|---|---|---|---|---|---|---|
| $(C_4N_2H_{14}Br)_4SnBr_6$ | 355 | 570 (530) | 105 (63) | 215 | 95±5 | 2.3 (2.1) |
| $(C_4N_2H_{14}I)_4SnI_6$ | 410 | 620 (626) | 118 (63) | 217 | 75±4 | 1.1 (1.1) |

$\lambda_{exc}$ is the wavelength at excitation maximum; $\lambda_{em}$ is the wavelength at the emission maximum, $\Phi$ is the photoluminescence quantum efficiency; $\tau_{av}$ is the photoluminescence lifetime; the values in parentheses are for 77K.

The broadband emissions with large Stokes shifts suggest that they are not from the direct excited states, but other lower energy excited states. In corrugated 2D and 1D metal halide perovskites, similar below gap broadband emissions have been observed as a result of exciton self-trapping.[11-14] It is well known for metal halides that the formation of localized self-trapped excited states is critically dependent on the dimensionality of the crystalline systems, and lowering the dimensionality would make exciton self trapping easier.[20-22] Therefore, 0D systems with strongest quantum confinement are reasonably expected to be favorable for the formation of self-trapped excited states. Indeed, the yellow emission from the 0D $(C_4N_2H_{14}Br)_4SnBr_6$ is very similar to the self-trapped 2.2 eV emission from $SnBr_2$ crystals at low temperature (12 K).[23] On the other hand, this class of 0D perovskites is similar in nature to a perfect host-dopant system with luminescent molecular species periodically embedded in an inert matrix without intermolecular interactions or band formation. Therefore, considering $SnX_6^{4-}$ as molecular species, the excited state processes for these 0D Sn halide perovskites can be depicted in the configuration coordinate diagram given in Fig. 2F. Upon photon absorption, the Sn halide octahedrons ($SnX_6^{4-}$) are excited to the high energy excited states, which undergoes ultrafast excited state structural reorganization to the lower energy excited states to generate strongly Stokes shifted broadband photoluminescence. Similar excited state structural reorganization, leading to strongly Stokes shifted broadband emissions, has been observed in a number of transition metal complexes, including Sn bromide complex [NEt$_4$]SnBr$_3$ in solutions.[24-28] Unlike corrugated 2D and 1D perovskites, with band formation due to the connections of metal halide octahedrons and structural distortion, emitting from both free exciton and self-trapped excited states at room temperature,[11-14] these 0D perovskites without band formation emit from the indirect reorganized excited states only. Indeed, these 0D perovskites allow us to unify the classic solid-state theory of "exciton self-trapping" with the molecular photophysics term of "excited state structural reorganization", as the building blocks (metal halide octahedrons) could be considered as either "crystal lattice points" or "molecular species". It should be pointed out that the 0D Sn halide perovskites present here are fundamentally different from the previously reported analogous compounds, such as Cs$_4$PbBr$_6$ and Cs$_2$SnI$_6$, which possess little-to-no quantum confinement for the individual metal halide octahedrons and exhibit emissions from the direct excited sates.[29-31] In other words, those previously reported 0D perovskites are not bulk assemblies of 0D core-shell quantum confined materials like the 0D Sn halide perovskites present here, but rather regular 3D crystalline materials.

The extremely high PLQEs in solid state make these earth-abundant lead-free perovskites highly promising for displays and solid-state lighting applications. Unlike conventional light emitters, such as organic emitters and colloidal quantum dots, requiring doping to prevent aggregation included self-quenching in solid state, these 0D perovskites are perfect host-dopant systems themselves. The strongly Stokes shifted broadband emissions without self-absorption are of particular interest for applications in down conversion white LEDs and luminescent solar concentrators. To demonstrate the application of these materials as phosphor, we fabricated down conversion LEDs, in which a commercial UV LED (340 nm) was used to optically pump polydimethylsiloxane (PDMS) films doped with ball-milled yellow emitting Sn bromide perovskite crystals and commercial blue emitting europium-doped barium magnesium aluminates ($BaMgAl_{10}O_{17}:Eu^{2+}$).[32] The UV LED (340 nm) was chosen considering the excitations of both the yellow and blue phosphors in the UV region (Extended Data Fig. 9). Fig. 3A shows the images of PDMS films doped with blue and yellow phosphors at different weight ratios under ambient light and a UV lamp irradiation. The emission spectra of UV pumped LEDs are shown in Fig. 3B. The CIE color coordinates and Correlated Color Temperatures (CCTs) are shown in Fig. 3C. A nice range of "warm" to "cold" white lights has been achieved by controlling the blending ratio between the two phosphors. With a blue/yellow weight ratio of 1:1, a decent white emission with CIE coordinates of (0.35, 0.39), a CCT of 4946 K, and a color-rendering index (CRI) of 70, was obtained. Excellent color stability was observed in this white LED at different operating currents, as shown in Fig. 3D. This could be attributed to the little-to-no energy transfer from the blue phosphors to the yellow phosphors, as there is a minimum overlap between the excitation of yellow phosphors and the emission of blue phosphors (Extended Data Fig. 9). The white LED also showed great stability in air with almost no change of light brightness and color during the preliminary testing, i.e. the device continuously on at ~ 400 $cd/m^2$ for more than six hours under the same operating power (Extended Data Fig. 10).

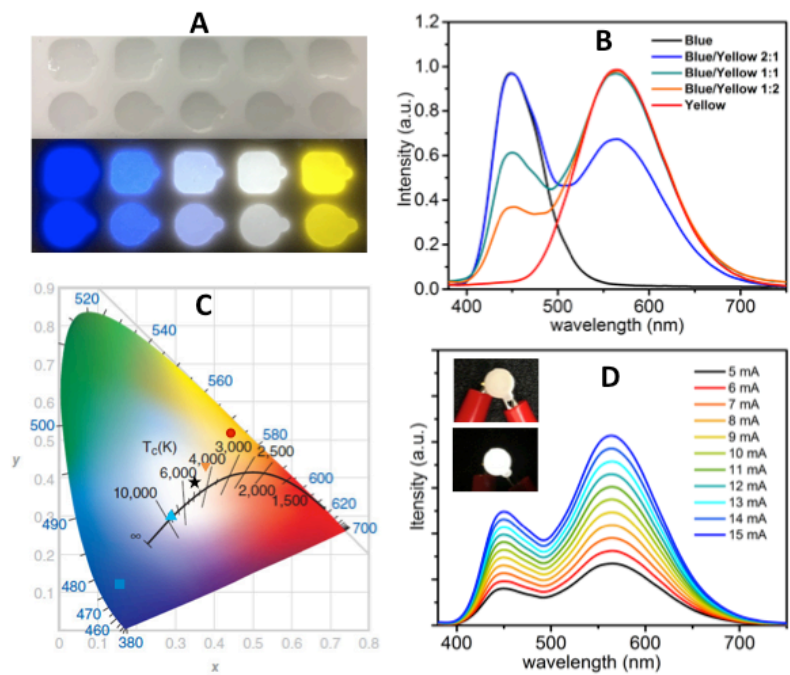

**Fig. 3. 0D Sn bromide perovskites as yellow phosphor for UV pumped white LEDs.** (A) Images of blue phosphors, yellow phosphors, and their blends with different weight rations (1:2, 1:1, and 2:1) embedded in PDMS under ambient light (top) and a hand held UV lamp irradiation (365 nm) (bottom). (B) Emission spectra of UV pumped LEDs with different blending ratios between blue and yellow phosphors. (C) CIE coordinates and CCTs for the UV pumped LEDs plotted on the CIE1931 chromaticity chart: blue (■), "cold" white (▲), white (★), "warm" white (▼), and yellow (●). (D) Emission spectra of a white LED at different driving currents, the insets show the device off and on.

By using appropriate organic and inorganic building blocks, we have been able to assemble a new class of organic-inorganic hybrid bulk quantum materials based on 0D lead-free tin halide perovskites, which represents a perfect host-dopant system with photoactive metal halide octahedrons periodically embedded in a wide band gap organic network through ionic bonds. Without band formation or quantum size effect, these ionically bonded bulk quantum materials enable access to the intrinsic properties of individual photoactive molecular species in bulk crystals, opening up new routes towards high performance photoactive crystalline materials for optoelectronic devices.

**Methods**

**Materials.** Tin (II) bromide, Tin (II) iodide, N, N'-dimethylethylenediamine (99%), γ-butyrolactone (GBL, ≥ 99%), hydrobromic acid (48 wt.% in H$_2$O), and hydriodic acid (55%) were purchased from Sigma-Aldrich. Dichloromethane (DCM, 99.9%), dimethylformamide (DMF, 99.8%), toluene (anhydrous, 99.8%), and ethyl ether (Stabilized with ~1ppm BHT) were purchased from VWR. Acetone (HPLC grade) was purchased from EMD Millipore. All reagents and solvents were used without further purification unless otherwise stated.

**N, N'-dimethylethylene-1,2- diammonium halide salts.** N, N'-dimethylethylene-1,2-diammonium bromide salts were prepared by adding hydrobromic acid solution (2.2 equiv) into the N, N'-dimethylethylenediamine (1 equiv) in ethanol at 0 °C. The organic salts were obtained after removal of the solvents and starting reagents under vacuum, followed by washing with ethyl ether. The salts were dried and kept in a desiccator for future use. N, N'-dimethylethylene-1,2-diammonium iodide salts were prepared following similar method.

**Solution growth of 0D Sn halide perovskite crystals.** Tin(II) bromide and N, N'-dimethylethylene-1,2- diammonium bromide were mixed at 1:4 molar ratio and dissolved in DMF to form a clear precursor solution. Bulk crystals were prepared by diffusing DCM into DMF solution at room temperature for overnight. The large colorless crystals were washed with acetone and dried under reduced pressure. The yield was calculated at ~ 70 %. C$_{16}$H$_{56}$N$_8$SnBr$_{10}$: Anal, Calc. C, 15.03; H, 4.42; N, 8.77. Found: C, 15.31; H, 4.24; N, 8.74. Tin(II) iodide and N, N'-dimethylethylene-1,2- diammonium iodide were mixed at 1:4 molar ratio and dissolved in GBL to form a clear precursor solution. Bulk crystals were prepared by diffusing DCM into GBL solution at room temperature for overnight. The large reddish crystals were washed with acetone and dried under reduced pressure. The yield was calculated at ~ 70 %. $C_{16}H_{56}N_8SnI_{10}$: Anal, Calc. C, 10.99; H, 3.23; N, 6.41. Found: C, 11.16; H, 3.23; N, 6.24.

**Single crystal X-ray diffraction (SCXRD).** Single crystal x-ray diffraction data of $(C_4N_2H_{14}Br)_4SnBr_6$ was collected using an Oxford-Diffraction Xcalibur-2 CCD diffractometer with graphite-monochromated Mo $K\alpha$ radiation. The crystal was mounted in a cryoloop under Paratone-N oil and cooled to 120 K with an Oxford-Diffraction Cryojet. A complete sphere of data was collected using $\omega$ scans with 1° frame widths to a resolution of 0.6 Å, equivalent to $2\theta \approx 72.5°$. Reflections were recorded, indexed and corrected for absorption using the Oxford-Diffraction CrysAlisPro software, and subsequent structure determination and refinement was carried out using CRYSTALS, employing Superflip to solve the crystal structure. The data did not allow for an unconstrained refinement: all hydrogens were restrained to the connecting nitrogen or carbon. The refinement was performed against $F^2$, with anisotropic thermal displacement parameters for all non-hydrogen atoms and with isotropic thermal displacement parameters for the hydrogens in the structure. $(C_4N_2H_{14}I)_4SnI_6$ was mounted on a nylon loop with the use of heavy oil. The sample was held at 100 K for data collection. The data were taken on a Bruker SMART APEX II diffractometer using a detector distance of 6 cm. The number of frames taken was 2400 using 0.3-degree omega scans with either 20 or 30 seconds of frame collection time. Integration was performed using the program SAINT which is part of the Bruker suite of programs. Absorption corrections were made using SADABS. XPREP was used to obtain an indication of the space group and the structure was typically solved by direct methods and refined by SHELXTL. The non-hydrogen atoms were refined anisotropically. VESTA was used as the crystal structure visualization software for the images presented in the manuscript.

**Powder X-ray diffraction (PXRD).** The PXRD analysis was performed on Panalytical X'PERT Pro Powder X-Ray Diffractometer using Copper X-ray tube (standard) radiation at a voltage of 40 kV and 40 mA, and X'Celerator RTMS detector. The diffraction pattern was scanned over the angular range of 5-50 degree (2θ) with a step size of 0.02, at room temperature. Simulated powder patterns were calculated by Mercury software using the crystallographic information file (CIF) from single-crystal x-ray experiment.

**Sn Nuclear Magnetic Resonance (NMR).** The $^{119}$Sn MAS NMR spectra were recorded on a Bruker Advance III HD spectrometer equipped with 4 mm MAS probe, operating at 186.5 MHz with the samples spinning at 12 kHz, high power proton decoupling, 30 s recycle delay, and typically 2048 scans. $SnO_2$ was used as a secondary reference at -604.3 ppm.

**X-ray photoelectron spectroscopy (XPS).** XPS measurements were carried out using a ULVACPHI, Inc., PHI 5000 VersaProbe II. The survey XPS spectra were recorded with a monochromatic Al Kα source using a 93.9 pass energy and 0.8 eV/step. High-resolution spectra were recorded using a 11.75 pass energy and 0.1/eV step. The high-resolution spectra binding energies were assigned using a C 1s binding energy of 286.2 eV for the C-N= bonds in the $(C_4N_2H_{14}Br)_4SnBr_6$. A binding energy of 487.0 eV for the Sn 3d5 was then found to correspond to that of Sn (II) in $SnBr_2$

**Absorption spectrum measurements.** Absorption spectra of both the bulk perovskite crystals were measured at room temperature through synchronous scan in an integrating sphere incorporated into the spectrofluorometer (FLS980, Edinburgh Instruments) while maintaining a 1 nm interval between the excitation and emission monochromators.

**Excitation spectrum measurements.** Excitation spectra of bulk perovskite crystals were measured at room temperature on a FLS980 spectrofluorometer (Edinburgh Instruments) monitored at maximum of emission spectra.

**Photoluminescence steady state studies.** Steady-state photoluminescence spectra of both bulk and microsize crystals in solid state were obtained at room temperature and 77 K (liquid nitrogen was used to cool the samples) on a FLS980 spectrofluorometer.

**Temperature dependent photoluminescence.** The temperature dependent photoluminescence spectra were measured on a Varian Cary Eclipse Fluorescence Spectrometer with a Water 4 Position Multicell Holder Accessory attached to a Julabo F12-EC Refrigerated/Heating Circulator filled with ethylene glycol-water mixture (3:2). It is interesting to see the PL intensity of 0D $(C_4N_2H_{14}Br)_4SnBr_6$ bulk crystals even increases a bit upon the increasing of temperature (recovering upon the decreasing of temperature), which is likely due to the change of refractive index of the bulk crystal samples (absorption increasing upon the increasing of temperature).

**Photoluminescence quantum efficiencies (PLQEs).** For photoluminescence quantum efficiency measurement, the samples were excited using light output from a housed 450 W Xe lamp passed through a single grating (1800 l/mm, 250 nm blaze) Czerny-Turner monochromator and finally a 5 nm bandwidth slit. Emission from the sample was passed through a single grating (1800 l/mm, 500 nm blaze) Czerny-Turner monochromator (5 nm bandwidth) and detected by a Peltier-cooled Hamamatsu R928 photomultiplier tube. The absolute quantum efficiencies were acquired using an integrating sphere incorporated

into the FLS980 spectrofluorometer. The PLQE was calculated by the equation:

$\eta_{QE} = I_S/(E_R - E_S)$, in which $I_S$ represents the luminescence emission spectrum of the sample, $E_R$ is the spectrum of the excitation light from the empty integrated sphere (without the sample), and $E_S$ is the excitation spectrum for exciting the sample. Control samples, rhodamine 101 and blue phosphor $BaMgAl_{10}O_{17}:Eu^{2+}$, were measured using this method to give PLQEs of ~ 98 % and ~ 93 %, which are close to the literature reported values. The PLQEs were double confirmed by a Hamamatsu C9920 system equipped with a xenon lamp, calibrated integrating sphere and model C10027 photonic multi-channel analyzer (PMA). The measurements taking account of indirect PL provided the same results within the error bars.

**Time-resolved photoluminescence.** Time-Resolved Emission data were collected at room temperature and 77 K (liquid nitrogen was used to cool the samples) using time-correlated single photon counting on a Horiba JY Fluoromax-4 Fluorometer. Samples were excited with 295 nm pulsed diode lasers. Emission counts were monitored at 530 nm. The average lifetime was obtained by multiexponential fitting.

**PL intensity dependence on excitation power density.** PL intensity versus power studies were carried out on an Edinburgh Instruments PL980-KS transient absorption spectrometer using a Continuum Nd:YAG laser (Surelite EX) pumping a Continuum Optical Parametric Oscillator (Horizon II OPO) to provide 360 nm 5 ns pulses at 1 Hz. The pump beam profile was carefully defined by using collimated laser pulses passed through an iris set to 5 mm diameter. Pulse intensity was monitored by a power meter (Ophir PE10BF-C) detecting the reflection from a beam splitter. The power meter and neutral density filters were calibrated using an identical power meter placed at the sample position. Neutral density filters and an external power attenuator were used to reduce the pump's power density to the desired power range. Detection consisted of an Andor intensified CCD (1024x256 element) camera collecting a spectrum from 287 nm to 868 nm and gated to optimize PL collection (typically a 30 to 50 ns gate depending on PL lifetime starting immediately following the 5 ns laser pulse). 100 collections were averaged at each power level with every laser pulse monitored to determine the average intensity. PL intensity was determined at the maximum of the PL emission curve.

**Materials photostability study.** To test the photostability, a 100 W 20 V mercury short arc lamp was used as continuous irradiation light source. The intensity of the irradiation was calibrated to 150 mW/cm$^2$. The emission was measured perodically on a HORIBA iHR320 spectrofluorimeter, equipped with a HORIBA Synapse CCD detection system.

**Thermogravimetry Analysis (TGA).** TGA was carried out using a TA instruments Q50 TGA system. The samples were heated from room temperature (around 22 °C) to 800 °C at a rate of 5 °C·min$^{-1}$, under a nitrogen flux of 100 mL·min$^{-1}$.

**UV pumped LEDs.** The blue (BaMgAl$_{10}$O$_{17}$:Eu$^{2+}$) and yellow ((C$_4$N$_2$H$_{14}$)$_4$SnBr$_{10}$) phosphors were blended with Sylgard 184 polydimethylsiloxane (PDMS) encapsulant, and put in a polytetrafluoroethylene (PTFE) mold to control shape and thickness. The whole mold was heated at 100 °C for 40 min in an oven to cure PDMS. The phosphors doped PDMS films were then attached to a UVTOP® UV LED with window, 340 nm, 0.33 mW (THORLABS) to form UV pumped LEDs. The LEDs were driven by a Keithly 2400 sourcemeter and emission spectra were recorded on an Ocean Optics USB4000 Miniature Fiber Optic Spectrometer. For device stability test, a white light LED was continuously powered by a Keithley 2400 at a stable current power to give a brightness of

~ 400 cd/m$^2$. Emission spectra were recorded at periodic intervals using an Ocean Optics USB4000 Miniature Fiber Optic Spectrometer.

**Acknowledgements** The authors acknowledge the Florida State University for financial support through the Energy and Materials Initiative. Jamie C. Wang acknowledges the National Science Foundation Graduate Research Fellowship under Grant No. DGE-1449440. The excitation power dependent luminescence measurements were performed on a transient absorption spectrometer supported by the National Science Foundation under Grant No. CHE-1531629. The authors also thank Dr. Lei Zhu for providing access to a fluorescence spectrophotometer and Dr. Hanwei Gao for providing access to the instrument for photostability test.

**Author Contributions** C.Z., Z.Y. and B.M. conceived the experiments and analyzed and interpreted the data. C.Z. and Z.Y. synthesized and characterized the materials; Y.T. performed PXRD, studied the photostability, and fabricated UV pumped LEDs; C.Z., J.W., K.H., Q. M. and P. D. measured the photophysical properties; E. H. and H.L. did XPS measurement, B. C. performed Sn NMR; R.C., J.N., T.B., and T.S. performed single crystal XRD analysis. The manuscript was mainly written by B.M. The project was planned, directed and supervised by B.M. All authors discussed the results and commented on the manuscript.

**Author Information** Reprints and permissions information is available at www.nature.com/reprints. The authors declare no competing financial interests. Correspondence and requests for materials should be addressed to B.M. (bma@fsu.edu).

**Extended Data Table 1 | Single crystal x-ray diffraction data and collection parameters.** The collection was performed at a temperature of ~ 120 K.

| Compound | $(C_4N_2H_{14}Br)_4SnBr_6$ | $(C_4N_2H_{14}I)_4SnI_6$ |
|---|---|---|
| Formula | $[(CH_3NH_2)_2C_2H_4]_4SnBr_{10}$ | $[(CH_3NH_2)_2C_2H_4]_4SnI_{10}$ |
| Molecular weight | 1278.40 g/mol | 1748.37 g/mol |
| Space group | $P$ -1 (# 2) | $P$ -1 |
| $a$ | 10.2070(4) Å | 10.7464(7) Å |
| $b$ | 10.6944(4) Å | 10.8924(7) Å |
| $c$ | 18.5996(6) Å | 11.1796(7) Å |
| $\alpha$ | 94.043(3)° | 64.2658(7) ° |
| $\beta$ | 102.847(3)° | 80.1825(7) ° |
| $\gamma$ | 97.904(3)° | 72.8331(7) ° |
| $V$ | 1949.89(12) Å$^3$ | 1124.94(12) Å$^3$ |
| $Z$ | 2 | 1 |
| $\rho_{calc.}$ | 2.177 g/cm$^3$ | 2.581 g/cm$^3$ |
| $\mu$ | 10.922 mm$^{-1}$ | 7.448 |
| Data collection range | 2.815° < $\theta$ < 34.220° | 1.986° < $\theta$ < 29.408° |
| Reflections collected | 57392 | 14019 |
| Independent reflections | 11532 | 5710 |
| Parameters refined | 540 | 165 |
| Restraints | 240 | 0 |
| $R_1$, w$R_2$ | 0.0651$^a$, 0.0511$^b$ | 0.0178, 0.0349 |
| Goodness-of-fit on $F^2$ | 0.9933 | 1.078 |

$^{a)}$ $R_1 = S \| F_o | - | F_c \| / S | F_o \|$. $^{b)}$ w$R_2 = [\Sigma w(F_o^2 - F_c^2)^2 / \Sigma w(F_o^2)^2]^{1/2}$.

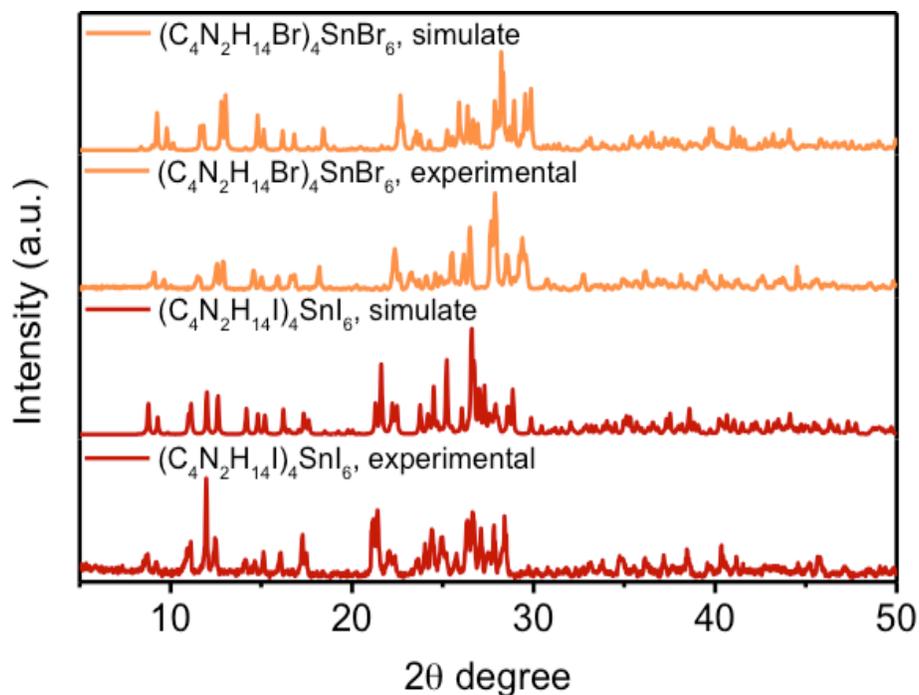

**Extended Data Figure 1 | PXRD of ground 0D tin halide perovskite bulk crystals as well as their simulated results.**

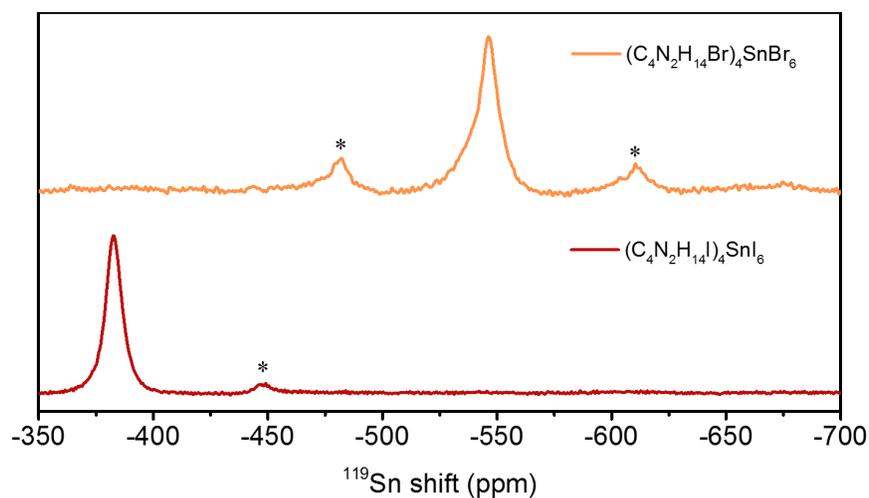

**Extended Data Figure 2 | $^{119}$Sn MAS NMR spectra of 0D Sn halide perovskites recorded at room temperature spinning at 12 kHz. Spinning sidebands are indicated with asterisks.**

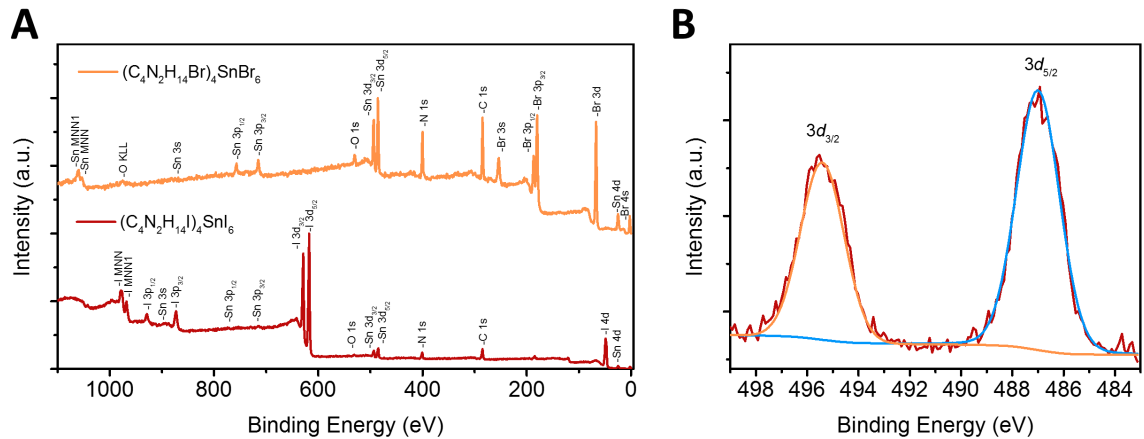

**Extended Data Figure 3 | X-ray photoelectron spectroscopy (XPS) of 0D tin halide perovskite bulk crystals: (a) survey spectra of two 0D perovskites, and (b) high resolution Sn spectra in 0D tin bromide perovskite.**

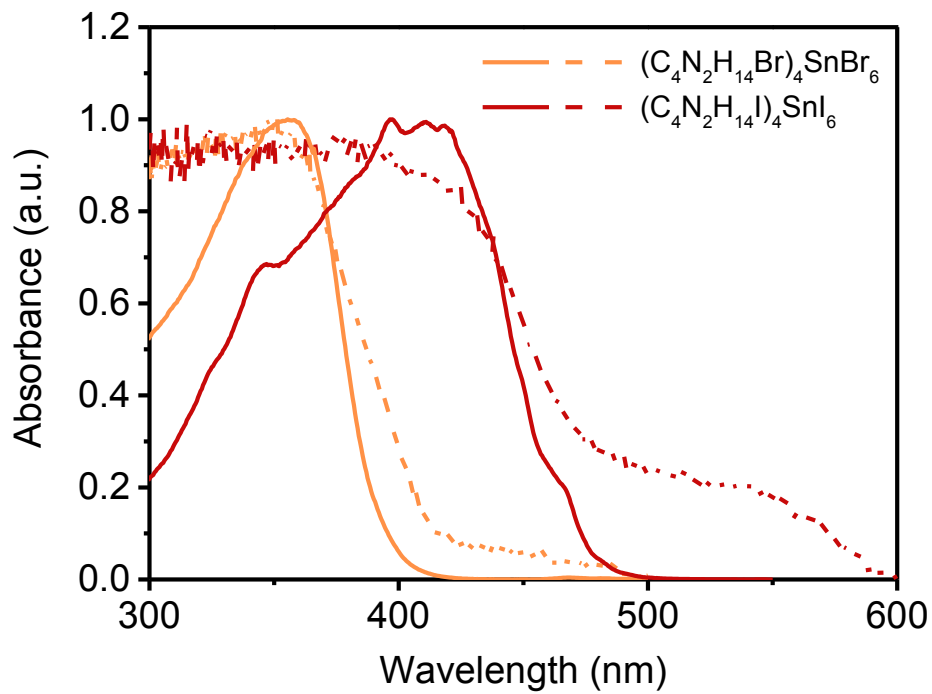

**Extended Data Figure 4 | Absorption (dash line) and excitation (solid line) spectra of 0D Sn halide perovskites recorded at room temperature.**

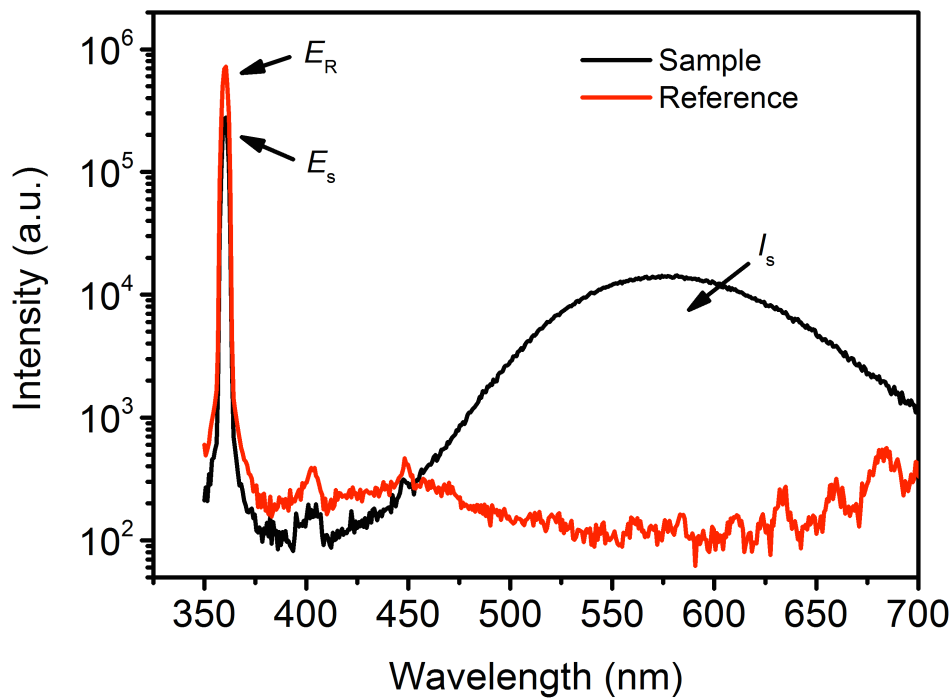

**Extended Data Figure 5 | Excitation line of reference and emission spectrum of 0D $(C_4N_2H_{14}Br)_4SnBr_6$ bulk crystals collected by an integrating sphere.** The PLQE was calculated by the equation: $\eta_{QE} = I_S/(E_R - E_S)$.

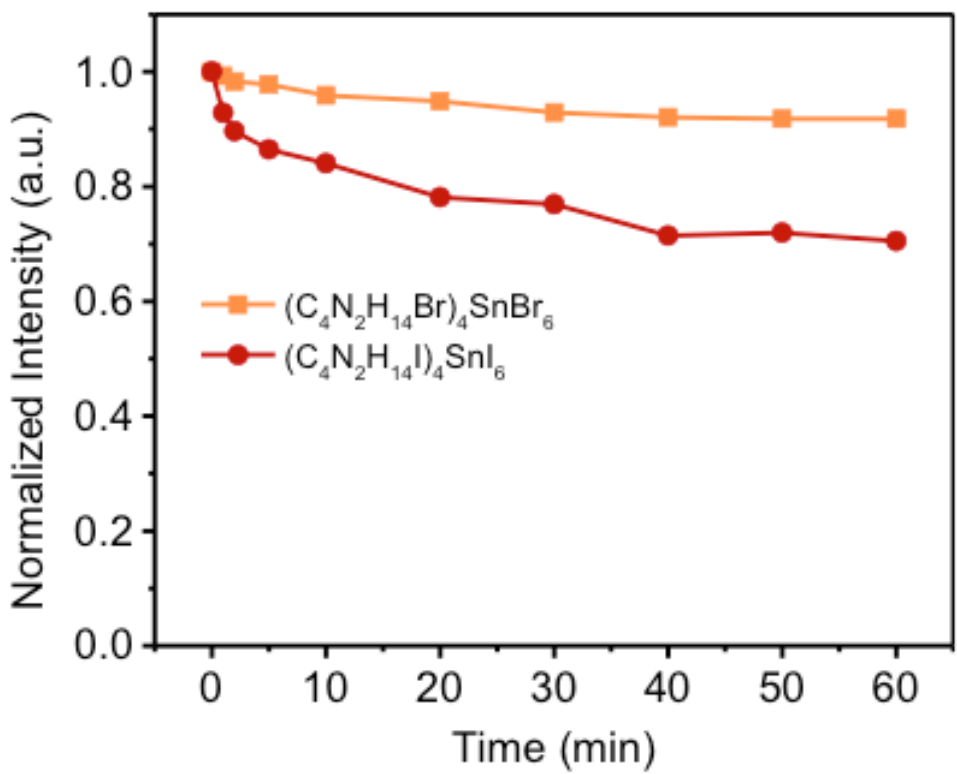

**Extended Data Figure 6 | Photostability of 0D Sn halide perovskites under continuous illumination using a high power mercury lamp (150 mW/cm$^2$).**

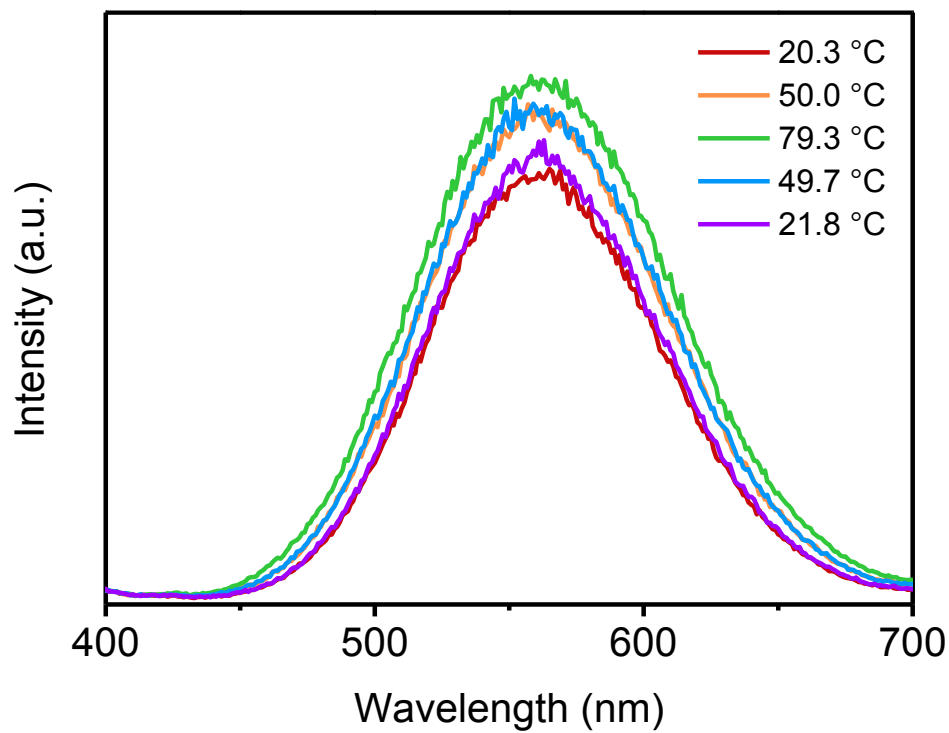

**Extended Data Figure 7 | Temperature dependent photoluminescence of 0D $(C_4N_2H_{14}Br)_4SnBr_6$.**

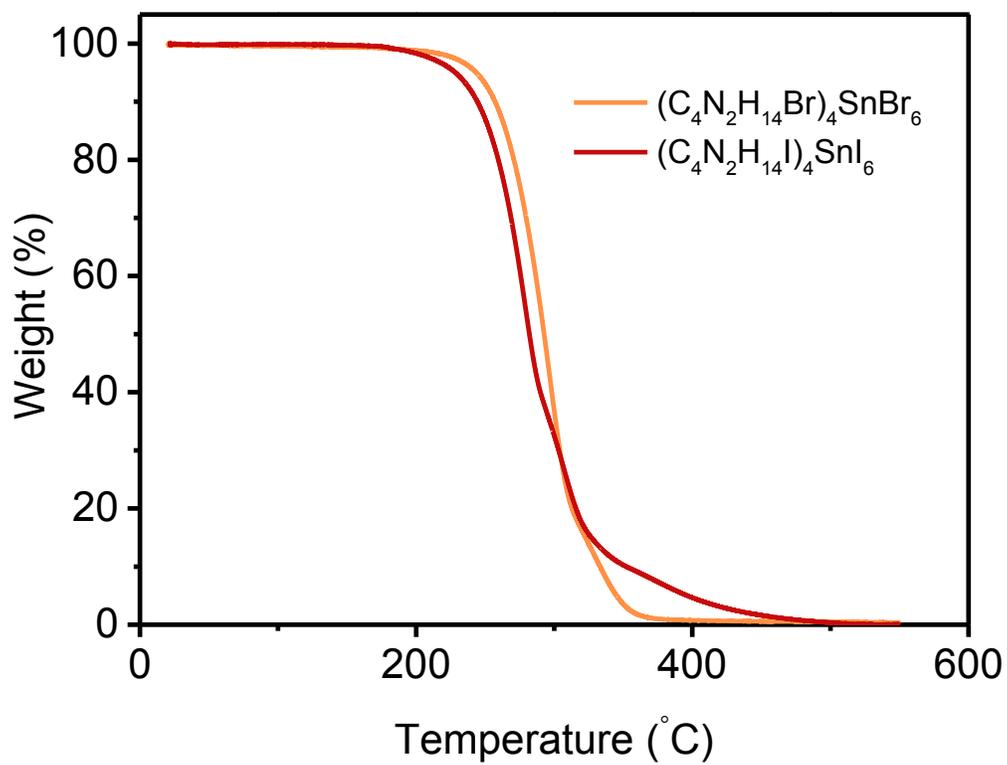

**Extended Data Figure 8 | TGA of 0D Sn halide perovskites**.

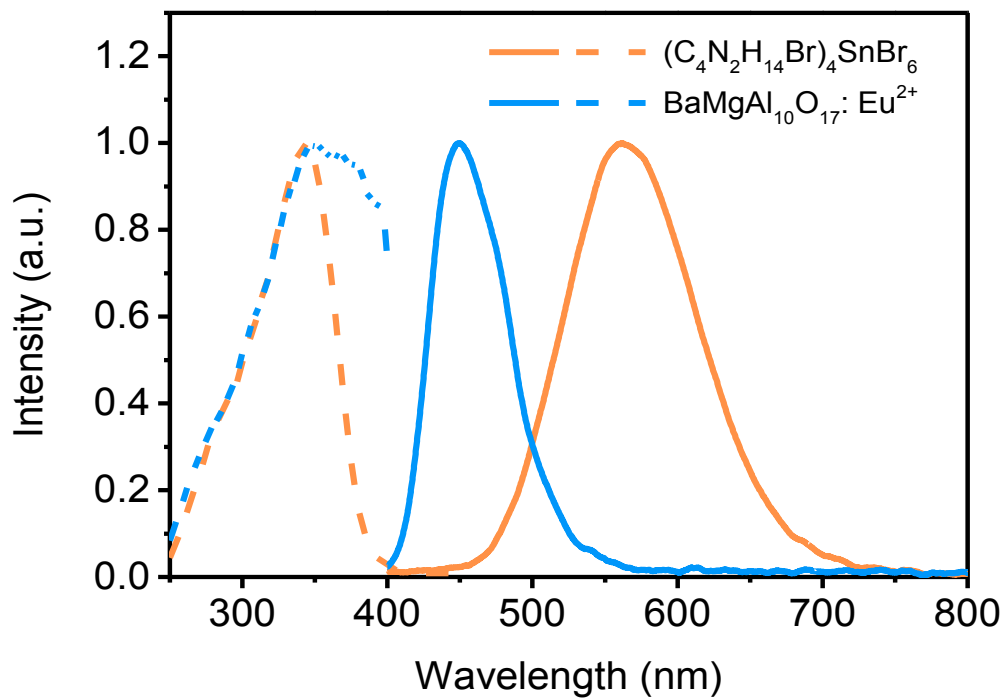

**Extended Data Figure 9** | The normalized excitation (dash lines) and emission (solid lines) spectra of $BaMgAl_{10}O_{17}:Eu^{2+}$ and $(C_4N_2H_{14}Br)_4SnBr_6$ phosphors.

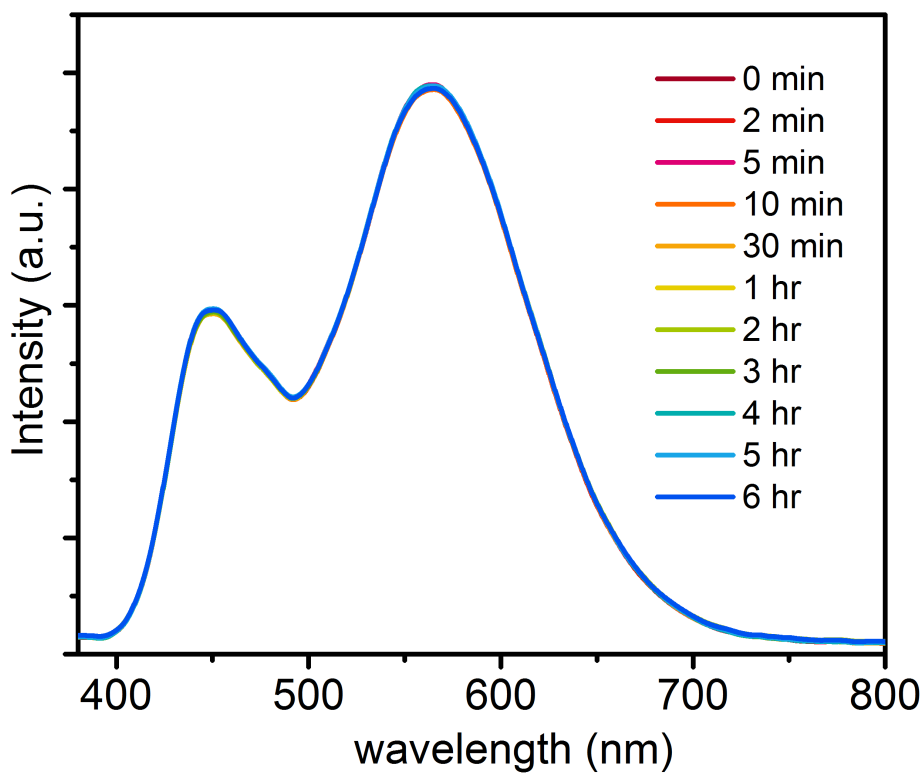

**Extended Data Figure 10** | Emission spectra of a white LED continuously operated in air for more than six hours with a brightness of ~ 400 cd/m$^2$.